# Influence of compressive strain on the hydrogen storage capabilities of graphene: A density functional theory study

*Vikram Mahamiya[a*], Alok Shukla[a*], Nandini Garg[b,c], Brahmananda Chakraborty[b,c*]*

[a]Indian Institute of Technology Bombay, Mumbai 400076, India

[b]High pressure and Synchrotron Radiation Physics Division, Bhabha Atomic Research Centre, Bombay, Mumbai, India-40085

[c]Homi Bhabha National Institute, Mumbai, India-400094

email: vikram.physics@iitb.ac.in; shukla@iitb.ac.in; brahma@barc.gov.in

**ABSTRACT**

Pristine graphene is not suitable for hydrogen storage at ambient conditions since it binds the hydrogen molecules only by van der Waals interactions. However, the adsorption energy of the hydrogen molecules can be improved by doping or decorating metal atoms on the graphene monolayer. The doping and decoration processes are challenging due to the oxygen interference in hydrogen adsorption and the clustering issue of metal atoms. To improve the hydrogen adsorption energy in pristine graphene, we have explored the hydrogen storage capabilities of graphene monolayer in the presence of compressive strain. We found that at 6 % of biaxial compressive strain, a 4*4*1 supercell of graphene can adsorb 10 $H_2$ molecules above the graphene surface. The average binding energy of $H_2$ for this configuration is found to be -0.42 eV/$H_2$, which is very suitable for reversible hydrogen adsorption. We propose that a 4*4*1 supercell of graphene can adsorb a total number of 20 $H_2$ molecules leading to a high hydrogen uptake of 9.4 %. The interaction between orbitals of carbon and hydrogen atoms and the charge transfer process have been studied by plotting the partial density of states and surface charge density plots. The electronic density around the C-C bonds of graphene increases in the presence of compressive strain, due to which hydrogen molecules are strongly adsorbed.

## 1. INTRODUCTION

Hydrogen energy is considered the most suitable alternative to fossil fuel since hydrogen is naturally abundant, environmentally friendly upon combustion, and possesses the highest energy per unit weight [1–3]. The use of fossil fuels generates hazardous gases like $CO_2$, CO, etc., putting health and lives at risk and polluting the environment [4,5]. To utilize hydrogen as a fuel, one has to store the hydrogen in a safe and compact manner. Also the method of storage should be affordable [6]. Hydrogen can be stored in gas, liquid, and solid-state form but the gas and liquid phase storage are not advisable because they require bulky pressure tanks and a high cost of liquefaction. The solid-state form is suitable for practical purposes provided that (1) the adsorption energy of $H_2$ should be in between -0.2 eV to -0.7 eV, and (2) hydrogen uptake of the system is more than 6.5 % following the guidelines of department of energy, united-states (DOE-US) [7].

In this regard, there are various substrates such as, metal alloys and hydrides [8–15], porous zeolites [16], metal-organic-frameworks [17–19], covalent triazine structures [20–23], carbon nanostructures [24-36], etc., have been explored for hydrogen storing purposes. Pristine carbon nanostructures bind $H_2$ molecules by small van der Waals interactions and, therefore, are not recommended for practical applications. So carbon nanostructures are decorated with various metals, including alkali metals, alkali-earth metals, and transition metals, and hydrogen molecules are adsorbed on these metal atoms. The strength of interaction increases by decorating the metal atoms because now additional electrostatic or Kubas type interactions are also present, which results in suitable hydrogen binding. There are some practical difficulties in the metal decoration which need to be addressed. Metal clustering is one of the challenges which can reduce the storage capacity drastically [36]. The other issue is the oxidation of metal atoms. Generally, the binding energy of oxygen on the metal atom is more as compared to hydrogen. Therefore, the oxygen interference can block the active adsorption sites for hydrogen, and the storage capacity gets reduced [37]. The adsorption energy of the attached $H_2$ can be improved by applying compressive strain to the substrate 2D structures. And therefore, sufficiently high uptake of hydrogen can be achieved without decorating the metal atoms.

Lamari and Levesque [38] reported that at 77 K and 1 MPa pressure, the hydrogen functionalized graphene can store up to 7 wt % of hydrogen, while at 293 K temperature and 30 MPa pressure the storage capacity is around 1.5 %. Wu et al. [39] have explored the effects of temperature, pressure, and geometry of three-dimensional pillared graphene on its hydrogen storage capacity by using molecular dynamics simulations. They report that the hydrogen uptake increases 3.76 fold when pressure is increased from 4 MPa to 15 MPa. Wang et al. [40] have reported 0.90 % of hydrogen uptake in graphene at 298 K temperature and 10MPa pressure.

Here, we have systematically investigated the influence of compressive strain on the average adsorption energy of the $H_2$ attached to the top of the graphene monolayer. We have plotted the partial density of states and surface charge density plots to understand the interaction mechanism and charge flow direction in presence of compressive strain.

## 2. METHODOLOGICAL DETAILS

Density functional theory (DFT) simulations are carried out by employing Perdew-Burke-Ernzerhof (PBE) [41] exchange-correlation functional as implemented in Vienna Ab Initio Simulation (VASP) package [42–45]. We have considered a 4*4*1 supercell of graphene for simulation purposes, and a vacuum of 20 Å is taken to avoid the periodic interactions along the z-direction. We have taken a Monkhorst-pack kpoints grid of length 5*5*1 to sample the Brillouin zone in DFT calculations. The plane wave basis expansion kinetic energy cutoff is 500 eV, while the convergence limit for the Hellman-Feynman force and energy are taken to be 0.01 eV/ Å and $10^{-5}$ eV, respectively. We have also included Grimme's DFT-D2 [46] dispersion corrections along with the PBE exchange-correlation to include the effects of the weak van der Waals interactions present in the system.

## 3. RESULTS AND DISCUSSIONS

The optimized structure of a 4*4*1 supercell of graphene with 32 carbon atoms is presented in **Fig. 1.** The lattice constants for this structure are, a = 9.84 Å and b = 8.53 Å. After obtaining the relaxed structure of graphene, we have kept hydrogen molecules on this structure and performed the relaxation calculations. A 4*4*1 supercell of graphene has ten complete hexagons, as shown in **Fig. 1.** Next, we have investigated the hydrogen adsorption properties of graphene by placing the first $H_2$ molecule at various sites of graphene, such as (1) above the center of the hexagon (C), (2) above the center of the bridge edge connecting two hexagons (B), and (3) above to the top of the carbon atom of graphene (T), as shown in **Fig. 1.**

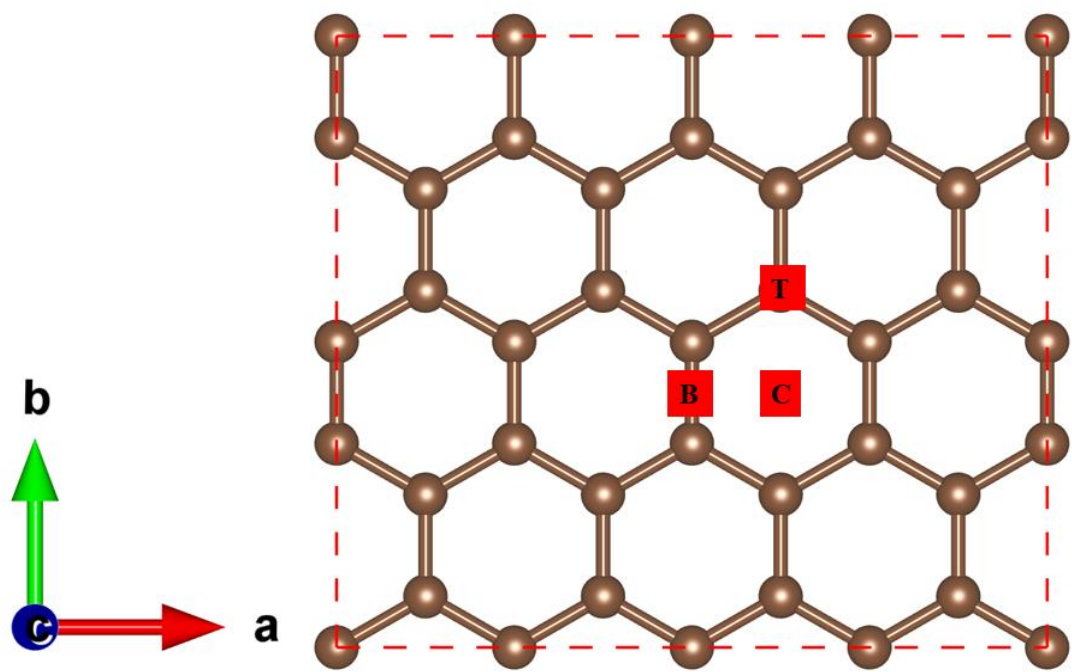

**Fig. 1. Optimized structure of a 4\*4\*1 supercell of graphene. Here, C, B, and T denotes top of the center of hexagon, bridge edge of the two hexagons, and top of the carbon atom of graphene sites, respectively.**

The hydrogen molecule is kept at around 2 Å distance away from the graphene sheet before the relaxation. To account the effect of the weak van der Waals interactions, we have corrected our GGA exchange-correlation level results by incorporating Grimme's DFT-D2 type dispersion corrections. The adsorption energy of the 1st $H_2$ kept at C, B, and T sites are found to be -0.11 eV, -0.10 eV, and -0.10 eV, respectively. Since the binding energy of the hydrogen molecule attached to the graphene does not lie in the criterion specified by DOE-US, pristine graphene is not suitable for hydrogen storage at ambient conditions. Another noteworthy point is that if we put one hydrogen molecule in front of the 4\*4\*1 supercell of graphene and one hydrogen molecule on the reverse side, the hydrogen uptake for the system will be 0.8 %, which is significantly lesser than the 6.5 %. Therefore, we have kept one hydrogen molecule on the top of each complete hexagon of **Fig. 1.** In this way, 10 $H_2$ molecules can be kept on the front side of graphene, and similarly, 10 H$_2$ molecules can be kept on the reverse side. The hydrogen uptake for such a system will be 9.4 %, which is sufficiently higher than the DOE limit of 6.5 %. Next, we have calculated the average adsorption energy of 10 H$_2$ molecules, which are kept on the top of the 10 hexagons of a 4*4*1 graphene monolayer. The average adsorption energy for 10 H$_2$ molecules is found to be -0.09 eV/H$_2$ computed using the GGA + DFT-D2 level of theory, which is not good enough for practical purposes because this adsorption energy will correspond to a desorption temperature lesser than room temperature. Cabria and Lopez [47] have reported 80-90 meV/H$_2$ average binding energy of hydrogen molecules attached on graphene and nanotube which is very similar to our study. Costanzo et al. [48] have reported 0.1 eV binding energy of the H$_2$ molecule attached on pristine graphene surface by accounting dispersion corrections using DFT-D scheme.

To improve the average adsorption energy of H$_2$, we have systematically applied compressive biaxial strain on the graphene supercell by reducing the lattice parameters of graphene. At 5 % of biaxial compressive strain, we report that the average adsorption energy of 10 H$_2$ is -0.29 eV/H$_2$, while it is -0.42 eV/H$_2$ at 6% of biaxial compressive strain. The average desorption temperatures of the 10 H$_2$ molecules attached to the graphene corresponding to binding energy -0.29 eV/H$_2$ and -0.42 eV/H$_2$, considering hydrogen molecules at ambient pressure, are found to be 371 K and 538 K, respectively. We have used the Van't Hoff formula [49] for the calculations of average desorption temperature given by the following equation:

$$\boldsymbol{T_d} = \left(\frac{\boldsymbol{E_b}}{\boldsymbol{k_B}}\right)\left(\frac{\Delta S}{R} - \boldsymbol{ln\ P}\right)^{-1} \qquad (\mathbf{1})$$

Here, $\boldsymbol{T_d}$ and $\boldsymbol{E_b}$ are the average desorption temperature and adsorption energy, respectively. $\boldsymbol{k_B}$, $\boldsymbol{\Delta s}$, $\boldsymbol{R}$, and $\boldsymbol{P}$ are Boltzmann constant, entropy change for H$_2$ while going into gas to liquid phase [50], gas constant, and atmospheric pressure, respectively.

This implies that if we apply around 5 % to 6 % of compressive strain to monolayer graphene (4*4 supercell), the average adsorption energy of hydrogen molecules falls in between -0.2 eV/$H_2$ to -0.4 eV/$H_2$, indicating that the $H_2$ adsorption process is reversible. The gravimetric weight percentage of hydrogen for this composition is 9.4 %, as described earlier. The optimized structure of graphene + 10 $H_2$ at 6 % of biaxial compressive strain is presented in **Fig. 2.**

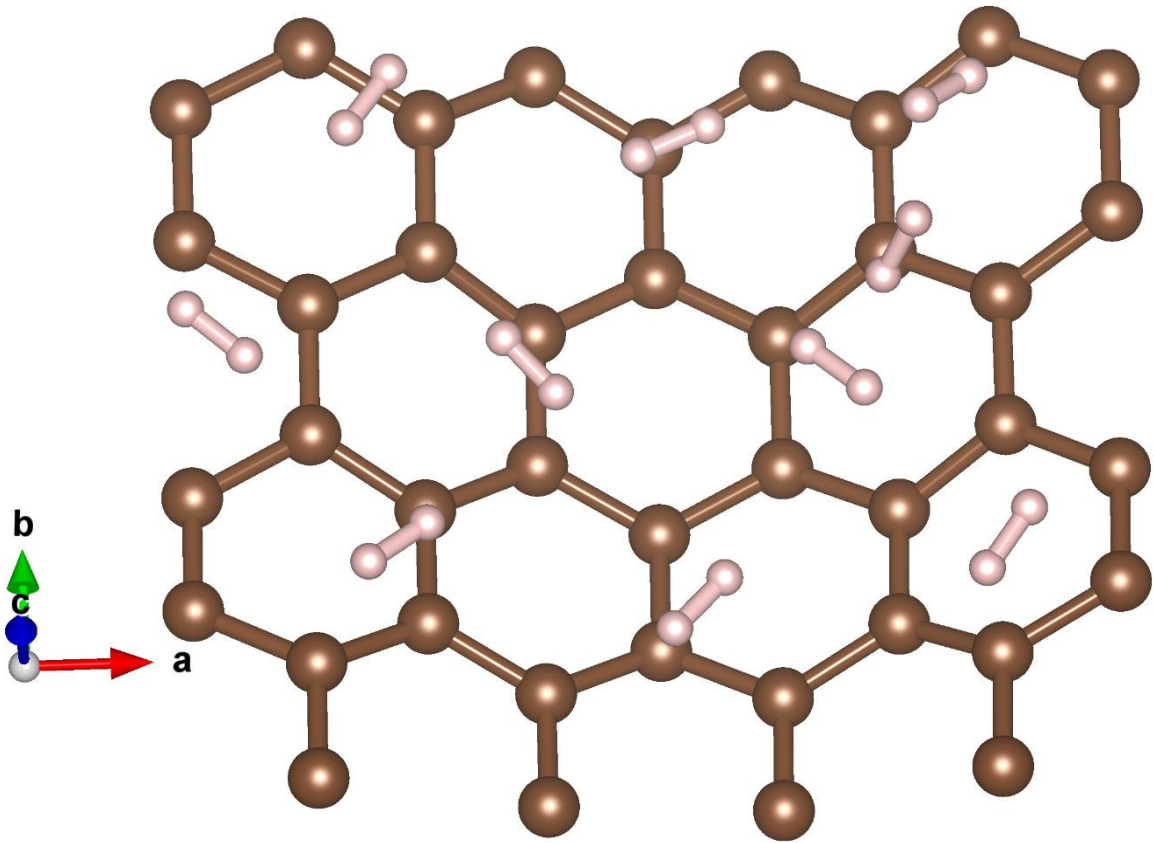

**Fig. 2. Optimized structure of graphene + 10 $H_2$ at 6% of biaxial compressive strain.**

The systematic improvement in the average adsorption energy of $H_2$ with the biaxial compressive strain is presented in **Table 1.**

**Table 1. Systematic improvement in the average adsorption energy of the hydrogen molecules attached to graphene monolayer by increasing the biaxial compressive strain.**

| Biaxial compressive strain (%) | Average binding energy of $H_2$ (eV/$H_2$) GGA + DFT-D2 |
|---|---|
| 0 | -0.09 |
| 2 | -0.12 |
| 3 | -0.13 |
| 4 | -0.19 |
| 5 | -0.29 |
| 6 | -0.42 |

Next, we have investigated the orbital charge flow mechanism between graphene monolayer and attached $H_2$ molecules by plotting the partial density of states of C 2(s & p) orbitals and H-1s orbital of graphene + 10 $H_2$ composition for unstrained graphene and graphene under 6 % of biaxial compressive strain as presented in **Fig. 3.** We notice that there is some enhancement in the states of C 2p orbitals near the Fermi level of the strained graphene compared to unstrained structure, as shown in **Fig. 3 (a & b).** Also, some depletion in states near the Fermi energy of H-1s orbital of strained graphene is observed compared to unstrained graphene, as shown in **Fig. 3 (c & d).** This implies that more charge transfers from H-1s orbital to C-2p orbitals of graphene + 10 $H_2$ composition takes place in the presence of compressive strain, which is responsible for the improvement in the binding energy of hydrogen molecules attached to the graphene monolayer.

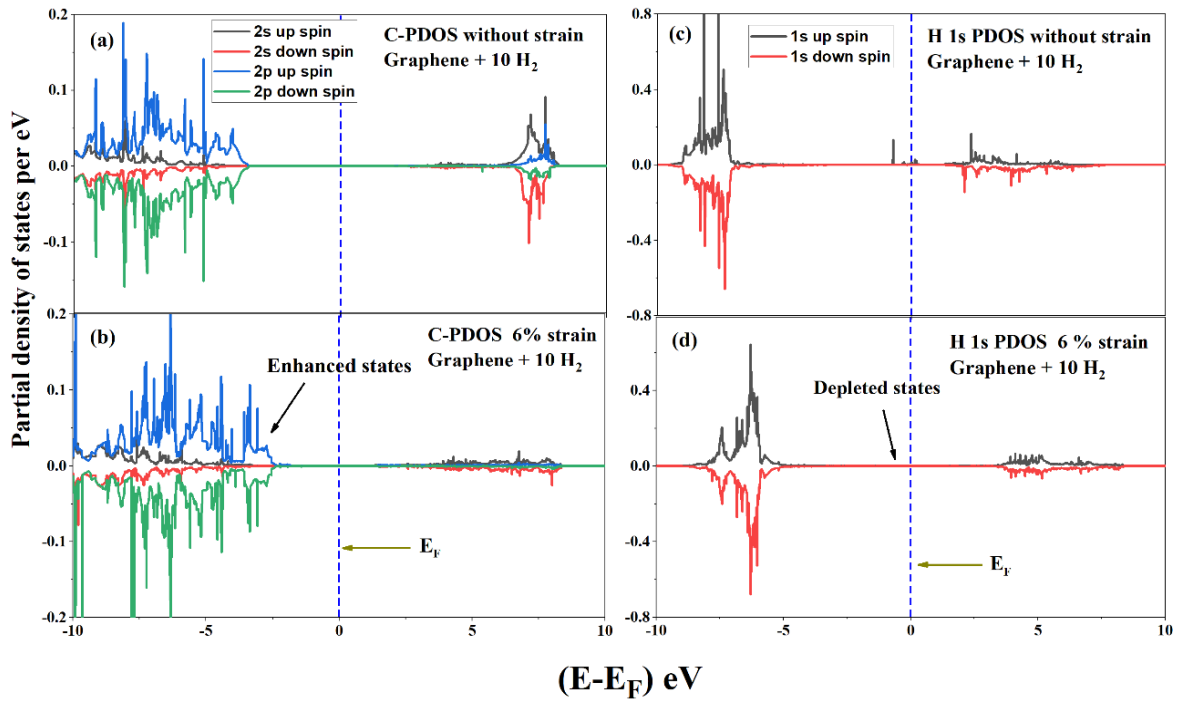

**Fig. 3. Partial density of states of (a) C-2 (s & p) orbitals for unstrained graphene + 10 $H_2$, (b) C-2 (s & p) orbitals of graphene + 10 $H_2$ at 6 % biaxial compressive strain, (c) H-1s orbital for unstrained graphene + 10 $H_2$, (d) H-1s orbital of graphene + 10 $H_2$ at 6 % of biaxial compressive strain.**

To verify the charge transfer process, we have plotted the surface charge density difference plots of graphene + 10 $H_2$ composition for pristine graphene and strained graphene with 6 % of compressive strain, as presented in **Fig. 4.** Here, **Fig. 4 (a)** denotes the charge density difference between ρ (graphene + 10 $H_2$) – ρ (graphene) for unstrained graphene, while **Fig. 4 (b)** denotes the charge density difference between ρ (graphene + 10 $H_2$) – ρ (graphene) for graphene under 6 % of compressive biaxial strain. These plots are according to B-G-R color pattern in which the red color around the hydrogen molecules denotes the charge loss region while the green and blue color around the carbon atoms of graphene denotes less and more charge accumulation regions, respectively. The isosurface value for **Fig. 4 (a & b)** is 0.080e.

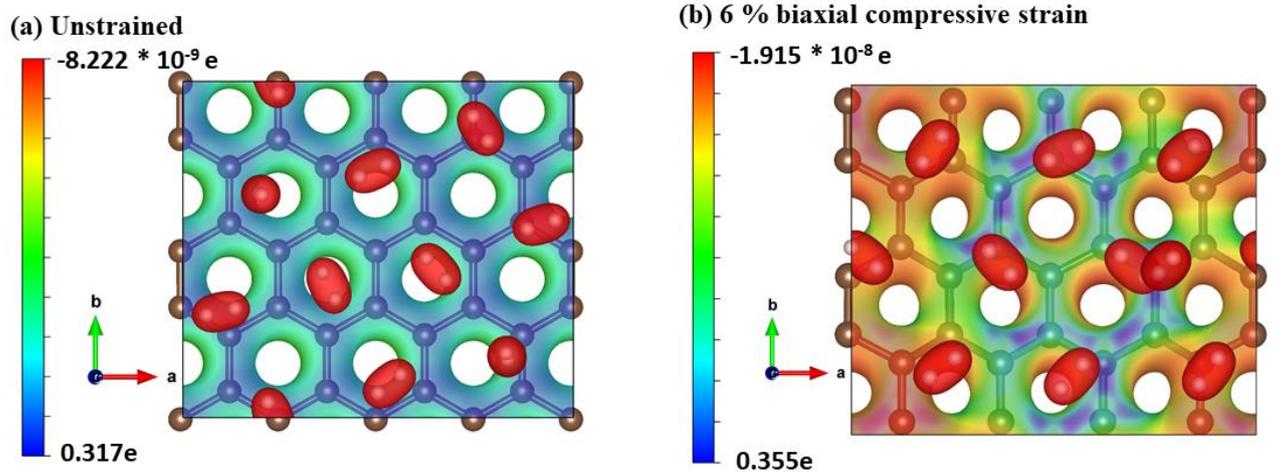

**Fig. 4. Surface charge density plots of ρ (Graphene + 10 H$_2$) – ρ (Graphene) (a) for unstrained composition (b) for strained composition with 6 % of biaxial compressive strain. The plots are according to B-G-R color pattern for isosurface value 0.080e.**

It is clear from **Fig. 4 (a),** that the charge has been uniformly distributed in graphene monolayer when H$_2$ molecules are attached in unstrained graphene structure, which gets redistributed between the carbon atoms of graphene at 6 % of biaxial compressive strain, as shown in **Fig. 4 (b)**. Also, the blue-colored region is more intense in the strained structure compared to unstrained structure, which implies that charge transfer from H$_2$ to graphene increases with compressive strain. This is one of the reasons behind the improvement of binding energy of hydrogen molecules attached to compressed graphene monolayer. Due to the systematic reduction of the lattice parameters with the increase in compressive biaxial strain, C-C bond lengths in graphene reduce, and, therefore, electron density around the C-C bonds increases which is also responsible for the strong binding of hydrogen molecules on the compressed graphene monolayer.

## 4. SUMMARY

In summary, we have investigated the hydrogen adsorption properties of graphene in the presence of compressive biaxial strain. We found that pristine graphene is not suitable for hydrogen storage, but the hydrogen adsorption energy can be systematically improved by applying uniform compressive strain to the graphene monolayer. We found that at 6 % of biaxial compressive strain, a 4*4*1 supercell of graphene monolayer can adsorb 10 $H_2$ molecules with an average adsorption energy of -0.42 eV/$H_2$, suitable for the reversible hydrogen storage. The hydrogen uptake for the system is found to be 9.4 % which is significantly higher than the DOE-US requirements of 6.5 %. Systematic reduction in lattice parameter results in the decrease in C-C bond lengths of graphene due to which electron density in C-C bond increases, leading to the strong binding of hydrogen molecules on the top of the graphene monolayer.

**ACKNOWLEDGEMENTS**

VM would like to acknowledge DST-INSPIRE for providing the senior research scholar fellowship and SpaceTime-2 supercomputing facility at IIT Bombay for the computing time.